# Strain engineering of magnetic states of vacancy-decorated hexagonal boron nitride

Bin Ouyang[1], Jun Song[1, a)]

1. Department of Mining and Materials Engineering, McGill University, H3A 0C5, Canada

Novel materials with tunable magnetic states play a significant role in the development of next-generation spintronic devices. In this paper, we examine the role of biaxial strain on the electronic properties of vacancy-decorated hexagonal boron nitride ($h$-BN) monolayers using density functional theory calculations. We found that the strain can lead to switching of the magnetic state for $h$-BN monolayers with boron vacancy or divacancy. Our findings promise a new route for the operation of low-dimensional spintronic devices.

The discovery of graphene[1] has spiked enormous research attention on low dimensional nanomaterials because of their fascinating properties and promising applications in nanoelectronic devices[1,2,3,4]. Among those one particular category of application is low-dimensional spintronics[5]. However, graphene-based materials often yield poor and unreliable control of the charge and magnetic states due to the zero band gap nature of graphene[6]. The hexagonal boron nitride ($h$-BN) monolayer, a two dimensional material with the same honeycomb structure as graphene, demonstrates many remarkable properties similar to graphene[7,8,9,10,11,12], and exhibits a wide direct band gap. The finite band gap avoids

---

a) The author to whom correspondence should be addressed. Email: jun.song2@mcgill.ca

coupling of conduction and valence bands[11], making *h*-BN monolayers more suitable for applications in spintronics than graphene[10,11,13]

Lattice defects, in particular vacancies, are often unavoidably present in *h*-BN monolayers as they are easily introduced during the fabrication or growth process[12,14]. These lattice imperfections constitute an important presence to enrich the properties of *h*-BN sheets[7,11,12,15,16,17]. For example, vacancies provide a source of magnesium to the otherwise magnesium-free *h*-BN system[11,15,17]. The new functionalities originated from lattice defects promise further applications for *h*-BN monolayers. Thus it is crucial to understand the mechanisms underlying those defect-induced functionalities and identify means to control them. Recently strain engineering has shown by several studies[9,18,19] to be an effective approach to tune properties in low dimensional naomaterials. For example, Choi et al.[19] reported that uniaxial strain can tailor confined states and collimation in graphene, and Guinea et al.[4] showed that strain aligned with certain crystallographic directions can induce strong gauge field to open energy gaps and create uniform local magnetic field in graphene. In addition, strains are also used to produce out-of-plane deformation (e.g., bending and rippling) to enable selective band engineering for low dimensional nanomaterials[3,20].

However, there are seldom studies[9,21] about the function of strain on *h*-BN monolayer, even less concerning the defective nature of it. In this letter, we examine the effects of biaxial strain on electronic and magnetic properties of *h*-BN monolayers defected by vacancies, through spin polarized first-principle calculations. We find that the strain can cause abrupt transition in the magnetic state for *h*-BN monolayers in the presence of certain vacancies, together with significant modifications of the electronic states and energy levels for the defective *h*-BN systems. These strain-induced behaviors can be quantitatively attributed to changes in the bonding structures and doping situations locally at the vacancy. Our findings explicitly demonstrate the interplay between mechanical strain and vacancies, and show strain engineering can provide an effective means to tailor magnetic state and regulate spin current in *h*-BN based nanoelectronic devices.

Spin polarized density-functional theory (DFT) calculations were performed using the Vienna ab-initio Simulation Package (VASP)[22] to study the structural, magnetic, and electronic properties of *h*-BN monolayers with vacancies. The vacancies considered include boron vacancy, nitrogen vacancy and divacancy, respectively abbreviated as B-V, N-V and Di-V in the following context for easy notations. Projector augmented wave method (PAW) and the general gradient approximation (GGA) are employed to deal with the exchange correlation functions. The Brillouin zone is sampled by $11\times11\times1$ and an energy cutoff of 500eV is used. In each calculation, a periodic simulation cell enclosing one vacancy (i.e., either a mono- or divacancy) is created, representing the situation of dilute vacancy concentrations. In the simulation cell, the interlayer separation is chosen as 14Å to eliminate the interactions between neighboring *h*-BN sheets across the periodic boundary. In particular, a $4\times4$ cell with 50 atoms is used for calculations for the system with a B-V or N-V, while a $6\times6$ cell with 72 atoms is used for the system with a Di-V. Calculations using larger cell sizes (i.e., up to $8\times8$ cell) are also performed, confirming no size dependence of our results.

An in-plane biaxial strain is imposed by varying the lattice constant *r* of the simulating cell and is computed as $(r-r_0)/r_0$ where $r_0$ is the lattice constant for a pristine stress-free *h*-BN monolayer. The strain is applied in two methods, in a discrete or step-wise fashion.

When the strain is applied in a discrete fashion, the monolayer is deformed directly from the stress-free state to the target strain value, followed by relaxation. When the strain is applied in a stepwise fashion, the monolayer is deformed from the stress-free state through a series of strain increments (*i.e.*, increment size of 1%) till the target strain is reached, during which the system is relaxed after each increment. For the range of strain examined, no difference is observed in the results for the two methods of applying strain. The biaxial strain considered in this study ranges from -10% to 10% and the properties of the system at individual strains are examined.

The evolutions of the total magnetic moment as functions of the applied strain are shown in Fig. 1 for systems with different types of vacancies. For the system with a B-V, the magnetic moment exhibits constant values of $1\,\mu_B$ or $3\,\mu_B$ when the strain is smaller than -5% or larger than -3% respectively, but undergoes an abrupt change in between. For the system with an N-V, the magnetic moment remains unchanged at $1\,\mu_B$ as the strain varies. For the system with a Di-V, the magnetic moment exhibits constant values of $0\,\mu_B$ or $2\,\mu_B$ when the strain is smaller than 1% or larger than 2% respectively but undergoes an abrupt change in between, a trend similar to the one observed in the case of B-V.

Along with the magnetic moment, the local bonding configuration around the vacancy also evolves as the strain varies, illustrated as inserted figures in Fig. 1 for each system. We can see that there is an apparent strain-induced transition in the local geometry at the vacancy for the systems with a B-V or Di-V, while the characteristics of the local geometry stay largely unchanged for the system with an N-V. To quantitatively elucidate the evolution of geometry at the vacancy during the straining of *h*-BN monolayers, we define a parameter *d/r* to quantify the local geometry change, with *d* being the shortest distance between two boron or nitrogen atoms that are neighboring the vacancy (in the case of Di-V, *d* is taken as the average of the shortest B-B and N-N distances) and *r* being the lattice constant previously defined. As plotted in Fig. 1d, the parameter *d/r* is a monotonically increasing function of the biaxial strain. We note from Fig. 1d that there are regimes of strains (denoted as transition regimes below) in which sudden jumps in the *d/r* value occur for systems with B-V or Di-V, while *d/r* rather evolves gradually as the strain varies for the system with an N-V. Specifically we note that the transition regimes coincide with the strain ranges where abrupt changes in the magnetic moment happen. Also worth noting is that the *d/r* value becomes larger than 1 for larger strains beyond those transition regimes.

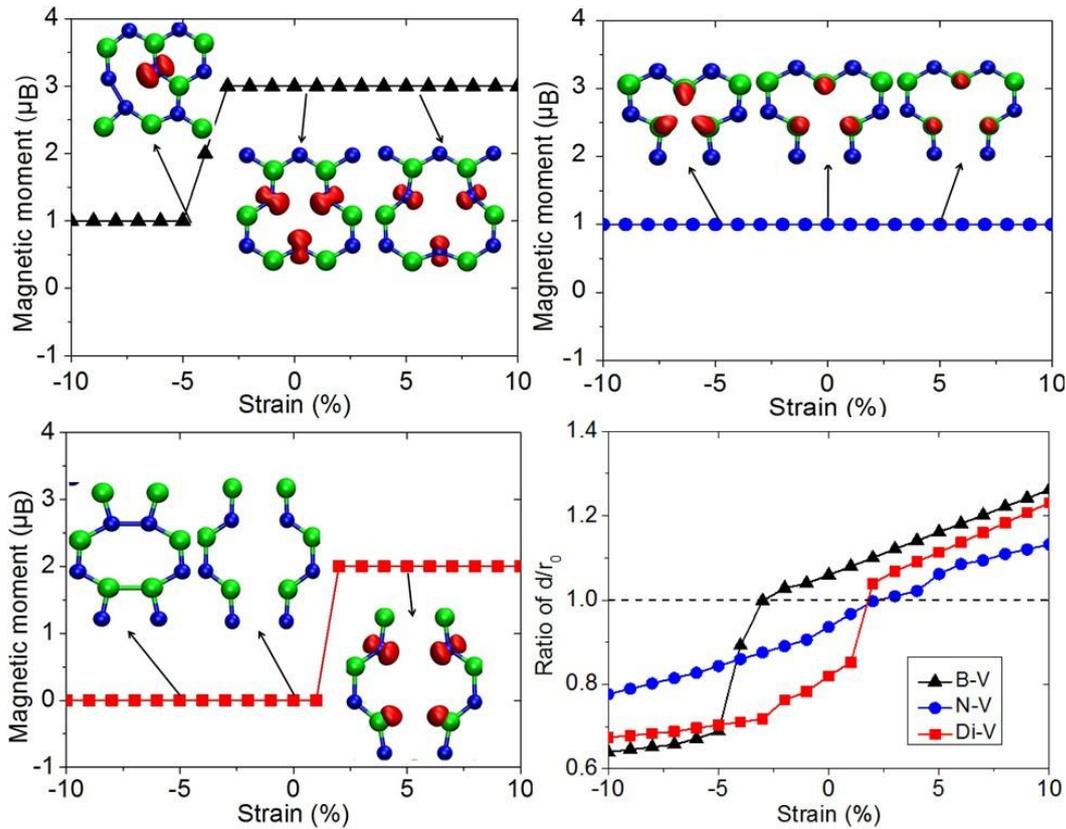

Fig. 1 (color online): The strain dependence of magnetic moment for systems with a (a) boron vacancy, (b) nitrogen vacancy and (c) divacancy. The local bonding configurations at the vacancy for three representative strains are shown as inserted figures for each system with the boron colored green and nitrogen atoms colored blue. The corresponding evolutions of d/r as functions of the strain are shown in (d) for different systems. The red isosurfaces in (a)-(c) represent the polarized spin at vacancies.

The coincidence of the change in magnetic moment and the jump in *d/r* suggests that the magnetic state stems from the local bonding configuration at the vacancy, the implication of which is discussed in details below. For the *h*-BN monolayer with a B-V, when it is subject to a strain less than -5%, the distance between one pair of then neighboring nitrogen atoms at the vacancy undergoes substantial shrinkage, inducing a Jahn-Teller distortion[23] to break the near three-fold symmetry as illustrated by the top left inserted picture in Fig. 1a. This leads to an intensive coupling between the *p* orbitals for the two nitrogen atoms within the pair. Consequently the two atoms completely lose the polarized spin that is otherwise present, leaving just one dangling bond. As a result the polarization is 1/3 of the one in the near three-fold symmetric configuration, and thus the magnetic moment becomes $1\mu_B$. On the other hand, the system roughly maintains a three-fold symmetry at the vacancy when the strain is larger than -3%. This local symmetry results in the presence of symmetric dangling $\sigma$ bonds and polarization of neighboring atoms[15] that produce a magnetic moment of $3\mu_B$ though limited Jahn-teller effect remains present[24]. The three-fold symmetry continues to hold as the strain increases because the increasing distance among dangling *p* orbitals does not lead to coupling but coulomb repulsion of dangling bonds. Consequently the separation between neighboring nitrogen atoms locally at the vacancy will be larger than the lattice constant to yield *d/r* > 1, consistent with what Fig. 1d shows.

In sharp contrast to the case of a B-V, the *h*-BN monolayer with an N-V however always maintains a near three-fold symmetry locally at the vacancy regardless of the applied strain as illustrated in Fig. 1b. The difference can be attributed to the charge transfer phenomenon between boron and nitrogen atoms in the *h*-BN system, shown in Fig. 2. We can see from Fig.

2 that the charge density around the N-V is much smaller than the one around the B-V, which is due to the different negativity of two types of atoms. As a consequence the coupling of *p* orbitals around the N-V is fairly weak and not sufficient to induce bond forming even under large compression strain. The coupling is apparently even weaker if the system is subject to tensile strain. Therefore the three-fold symmetry is always retained in the N-V defected *h*-BN system and the magnetic moment remains constant at $1\,\mu_B$ with no change in the overall magnetic configuration (cf. Fig. 1b) during straining.

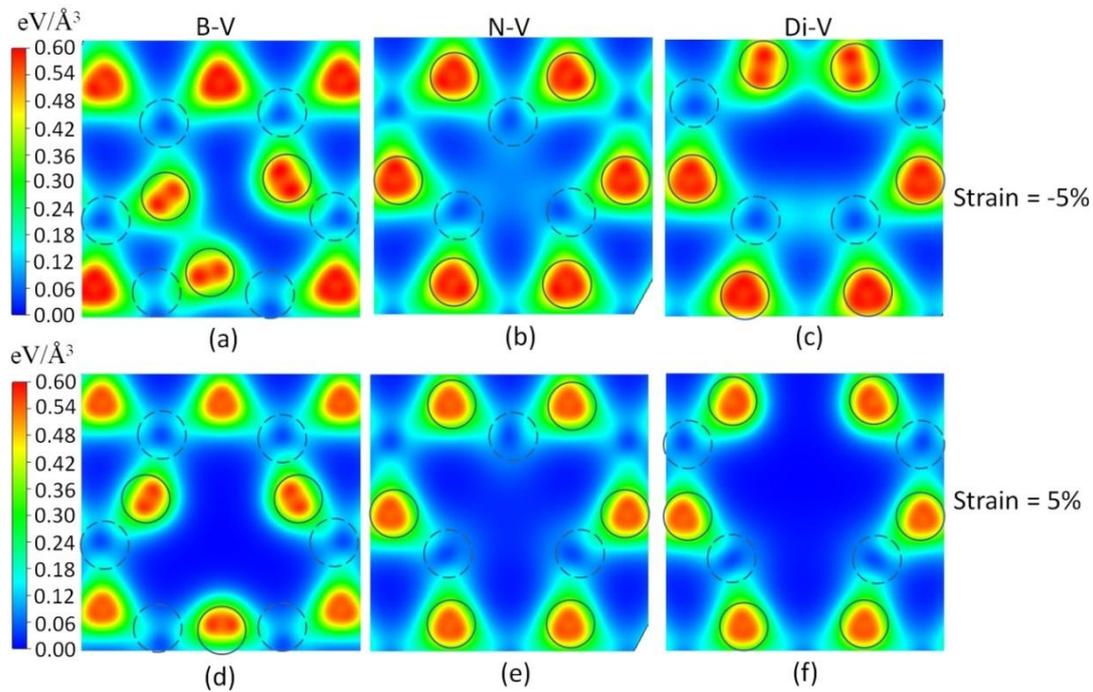

Fig. 2 (color online): Contour plots of charge density in the h-BN monolayer around (a) boron vacancy, (b) nitrogen vacancy, (c) divacancy under -5% biaxial strain, and (d) boron vacancy, (e) nitrogen vacancy, (f) divacancy under 5% biaxial strain. The dashed and solid circles indicate the position of boron and nitrogen atoms neighboring the vacancy respectively.

For the case of a Di-V, the atoms form a stable 5-8-5 loop structure at the vacancy when the strain is less than 1%, as shown in Fig. 1c. The 5-8-5 loop structure leaves no dangling bond[17,25] in the system and thus a magnetic moment of $0\,\mu_B$ is expected. However, this loop structure is broken when the system is strained beyond 2%, leading to a transition in the local geometry at Di-V that is indicated by the evolution of *d/r* in Fig. 1d. Following this transition, the system is no longer magnetism-free and exhibits a magnetic moment of $2\mu_B$. The transition is also clearly visible from the charge density plots shown in Figs 2c and 2f, evidenced by the disappearance of strong coupling (i.e., overlapping) between the charge density contours within the B-B and N-N pairs immediately neighboring the Di-V after the breakage of the 5-8-5 loop. To understand the origin of the magnetism, we examine the local atomic details at the Di-V. We see from Figure 1c that the breakage of the 5-8-5 loop structure creates dangling bonds and polarized spin. As Si et al.[15] pointed out, the magnetic moment mainly comes from (i) the atoms with dangling bonds and (ii) other polarized atoms in the close vicinity of the vacancy, where the polarization of (ii) is often derived from (i). Comparing Figs 2d-f, we can see that the local electron structure at Di-V includes the characteristics from the ones at both boron and nitrogen vacancies. In this regard, the magnetic state of the Di-V can be roughly viewed as a combination (yet of half the magnitude) of the ones of B-V and N-V[11] and thus a magnetic moment of $2\mu_B$ is expected[26].

From above we see that the biaxial strain can induce transformation in the local geometry at the vacancy to change the magnetic state of the system. Since the equilibrium bonding configuration is presumably determined by the distribution and hybridization of energy levels, we compute the corresponding spin-polarized band diagrams (supplemented by spin polarized density of states calculations in Supplemental Material), illustrated in Fig. 3 for systems with different vacancies at three representative strains of -5%, 0 and 5%. For the system with a B-V, it exhibits p-type doping as the nitrogen atoms with dangling p orbitals neighboring the vacancy act as triple acceptors. When the system is subject to biaxial straining, we see that all the band levels shift downwards together with the Fermi level decreasing as the strain increases. In addition we see that two degenerate doping levels appear along with a considerable change in the Fermi level when a large compressive strain is applied (see Fig. 3a and Supplemental Material). The formation of degenerate energy states is also the reason underlying the strong Jahn-Teller distortion and the abrupt change in magnetic moment shown in Fig. 1a. The system with an N-V on the other hand exhibits n-type doping because the boron atoms neighboring the vacancy act as donors of electrons[15]. We see from Fig. 3 that with the increasing of strain, similar to the case of B-V all the band levels move downwards, but the Fermi moves up instead. As the strain increases, the conduction band and the doping level approach each other, and eventually overlaps to induce hybridization of energy levels (cf. Fig. 3f). In contrast to the case of B-V, the system with an N-V does not show the formation of degenerate states as strain varies, which is consistent with the invariance of its magnetic state as shown in Fig. 1b. For the system with a Di-V, we again observe that all the band levels move downwards as the strain increases. Also we note that the system undergoes a transition from no spin polarization (e.g., Figs 3g-h at strains of -5% and 0%) to non-zero spin polarization (e.g., Fig. 3i at strain of 5%) as the strain increases. The strain-induced transition in spin polarization directly reflects the evolution of the magnetic moment previously shown in Fig. 1c. Meanwhile gaps between degenerate levels in different spin states are observed when the system exhibits spin polarization (e.g., see Fig. 3i). Both the polarization and splitting of energy levels are found to be direct consequences from the breakage of the 5-8-5 loop, during which one B-B and one N-N bonds immediately neighboring the divacancy are broken. As a result, a mixture of n-type and p-type doping levels are expected due to the n-type nature of dangling boron atoms and the p-type nature of dangling nitrogen atoms.

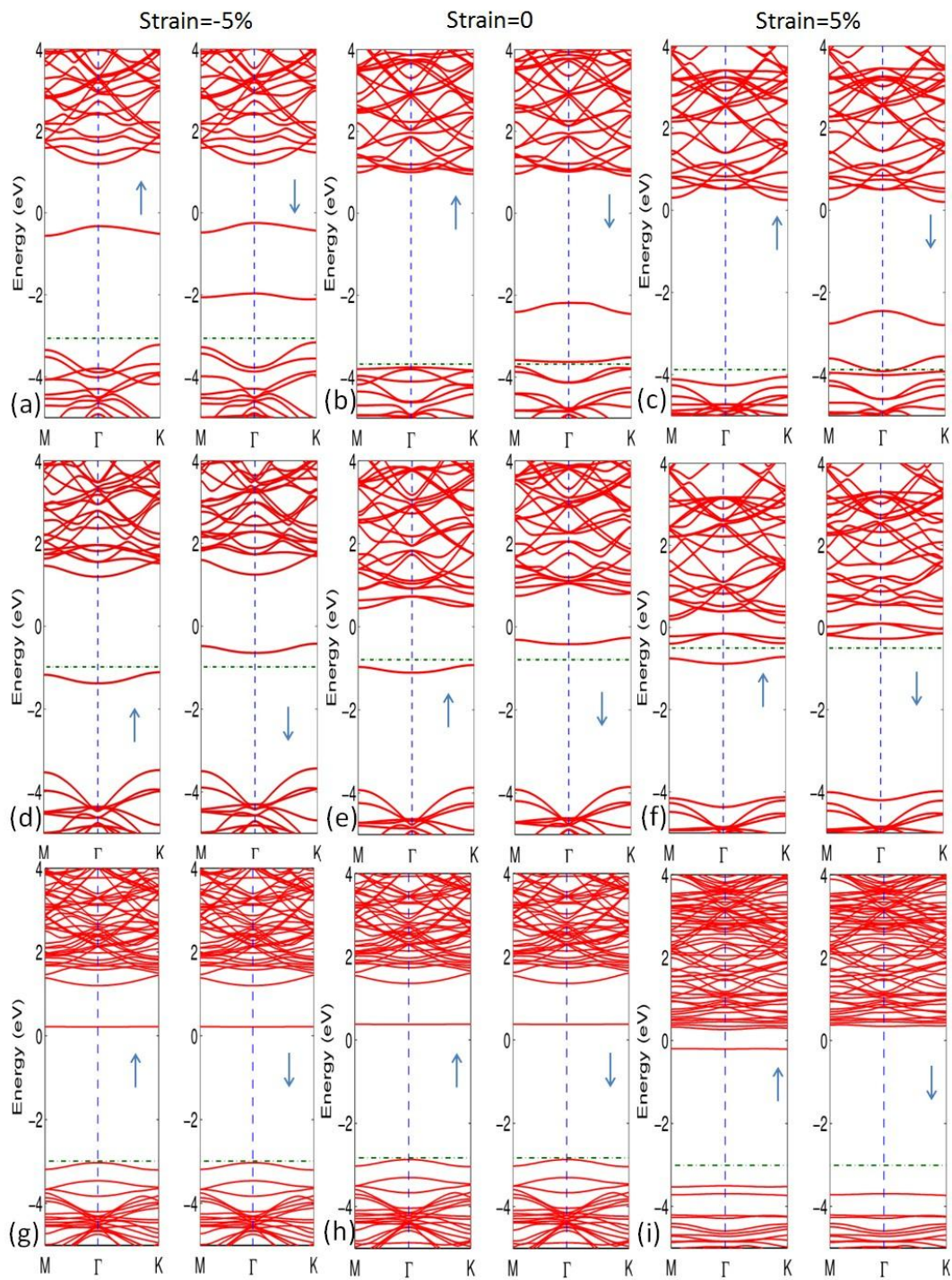

Fig. 3 (Color online): Band structures for h-BN systems with a boron vacancy (a)-(c), nitrogen vacancy (d)-(f), and divacancy (g)-(i) under different biaxial strains. The figures on the left, middle and right columns correspond to the band structures calculated at the representative strains of -5%, 0% and 5% respectively. The vertical dashed line indicates the Brillouin boundary at gamma point while the horizontal dotted line indicates the Fermi level.

In summary, strain engineering of vacancy-decorated *h*-BN monolayers has been examined by DFT calculations. Our study shows that the magnetization induced by the presence of vacancies can be tuned via biaxial straining. In particular, we show that the straining, though has no effects on the magnetic state of the system with a nitrogen vacancy, can induce abrupt transitions in the magnetic moment of systems with a boron vacancy or divacancy, for the range of strains examined. The changes in the magnetic moment are found to be directly correlated with the transformations in the local bonding configuration at the vacancy and the formation of degenerate energy levels. Furthermore the transitions in the magnetic moment, when occurred, are observed to take place within a fairly narrow regime of strains, with the magnetic moment being invariant w.r.t the strain prior to or after the transition, *i.e.*, the strain can act as an effective *switch* to "flip" the vacancy-decorated *h*-BN system between discrete magnetic states. Considering that the magnetization switching is often closely accompanied by changes in electrical transport properties and spin-dependent barriers[27,28], this promises new route towards the operation of low-dimensional spintronics (e.g., regulate spin current in spin-based field-effect transistors[28,29]) by strain instead of electric field. In addition to the magnetic state, we also show that the band gap and doping levels of the vacancy-decorated *h*-BN system can be modified by the biaxial strain, which provides important reference for the application of *h*-BN monolayers as building blocks in p-n junctions or heterostructures.

**Acknowledgements:**

The authors acknowledge support of this work by the NSERC Discovery grant (grant # RGPIN 418469-2012). The authors also would like to acknowledge Supercomputer Consortium Laval UQAM McGill and Eastern Quebec for providing computing power.